\documentclass[smallextended]{svjour3}       
\smartqed  

\usepackage{fullpage}
\AtBeginDocument{%
  \providecommand\BibTeX{{%
    \normalfont B\kern-0.5em{\scshape i\kern-0.25em b}\kern-0.8em\TeX}}}
\usepackage[numbers,sort&compress]{natbib}

\usepackage{balance}
\usepackage[utf8]{inputenc}
\usepackage[T1]{fontenc}
\usepackage{amsmath}
\usepackage{algorithmic}
\usepackage{graphicx}
\usepackage{hyperref}
\usepackage{textcomp}
\usepackage{xcolor}
\usepackage{todonotes}
\usepackage{fancyhdr}
\usepackage{tcolorbox}
\usepackage{multirow}
\usepackage{threeparttable}
\usepackage{booktabs}

\newcommand{\revised}[1]{#1}
\newcommand{\reviseX}[1]{\textcolor{black}{#1}}
\usepackage{xspace}
\newcommand{\lang}{\textit{language}\xspace}
\newcommand{\experimental}{\textit{experimental/personal}\xspace}
\newcommand{\excl}[1]{}

\begin{document}
\title{Development and Evolution of Xtext-based DSLs on GitHub: An Empirical Investigation}

\author{Weixing Zhang         \and
        Daniel Strüber \and
        Regina Hebig
}


\institute{Weixing Zhang \at
              Chalmers $|$ University of Gothenburg, Gothenburg, SE \\
              \email{weixing.zhang@gu.se}           
            \and
            Daniel Strüber \at
              Chalmers $|$ University of Gothenburg, Gothenburg, SE\\
              Radboud University, Nijmegen, NL \\
              \email{danstru@chalmers.se} 
            \and
           Regina Hebig \at
              University of Rostock, Rostock, DE \\
              \email{regina.hebig@uni-rostock.de}
}

\date{Received: date / Accepted: date}

\maketitle

\begin{abstract}
Domain-specific languages (DSLs) play a crucial role in facilitating a wide range of software development activities in the context of model-driven engineering (MDE). However, there exists a significant gap in the systematic understanding of how DSLs evolve over time, which could hamper the development of effective methodologies and tools. To address this gap, we performed a comprehensive investigation into the development and evolution of textual DSLs created with Xtext, a particularly widely used language workbench in the MDE.
We systematically identified and analyzed 1002 GitHub repositories containing Xtext-related projects.
A manual classification of the repositories brought forward 226  ones that contain a fully developed language.
We further categorized the latter into 18 separate categories of application domains, studied their contained DSL definition artifacts and analyzed the extent to which example instances using the grammar are available.
In addition, we explored DSL development practices, focusing on the development scenarios involved, evolution activities, and the modification and co-evolution of related artifacts.
We observed that DSLs are used more, evolve faster, and are maintained longer in specific domains, such as Data Management and Databases. We found grammar definitions of DSLs in 722 repositories in total, but only about a third of them provided corresponding textual instances, and most of them used more than 60\% of the grammar rules. Considering different language development approaches, we found that the majority of analyzed languages were developed following a grammar-driven approach, although a notable number adopted a metamodel-driven approach. Additionally, we identify a trend of retrofitting existing languages in Xtext, illustrating the framework's flexibility beyond the creation of new DSLs. By investigating software evolution aspects, we found that the development lifecycle of DSLs varies, but in most DSL development projects, updates to grammar definitions and example instances are very frequent, and most of the evolution activities can be classified as ``perfective'' changes.
Addressing a need for large and systematically documented datasets in the model-driven engineering community, we contribute a dataset of  repositories together with our collected meta-information, which can be used to inform the development of improved tools for supporting the development and evolution of DSLs.
\end{abstract}



\section{Introduction}
Domain-specific languages (DSLs, \cite{kosar2016domain}) are custom-tailored software languages addressing a particular domain of expertise.
By providing a tool to create models on a suitable abstraction level, DSLs play an important role in model-driven engineering (MDE, \cite{stahl2006model}), where models created using a DSL can be used for a large variety of activities such as design, analysis, code generation,  and testing.

Developing a DSL is a high-stakes activity.
\excl{Any decision made during the design of a DSL affects its usefulness as well as the way it is used.}
Previous design decisions often cannot be changed without significant effort on the part of the language developers and users.
Still, a need to change the language may arise especially in the context of \textit{language evolution}, where the developers add new features or respond to experience with the language \cite{lammel2018software}.
In consequence, there is a need for sound methods, practices, and techniques for supporting the evolution of DSLs.
However, to date, the development of support for DSL evolution is typically driven by the opinion of experts and individual cases encountered in their own practice or experience reports---Thanhofer et al. \cite{thanhofer2017systematic} provide a survey with 14 individual cases.
Developers of future evolution methods would benefit from systematic knowledge about DSL evolution obtained from a larger number of cases.

To understand how MDE artifacts are developed and evolved, there is a trend towards large-scale studies that systematically collect evidence from open-source software (OSS) projects \cite{Hebig2016mining}~\cite{shrestha2023evosl}~\cite{babur2024language}. \excl{Examples are studies of UML models at {MODELS'16}~\cite{Hebig2016mining}, on the evolution history of Simulink models at {MODELS'23}~\cite{shrestha2023evosl}, and on the usage of EMF concepts in {EMSE'24}~\cite{babur2024language},}
However, for development of DSLs in the context of MDE, such a study is not available yet.

In this paper, towards closing this gap, we contribute the first large-scale multiple-case study of DSL development and evolution.
Based on a repository mining methodology \cite{kalliamvakou2014promises}, we collected and analysed data from 1002 GitHub repositories, to answer questions about involved artifact types, development scenarios, and evolution activities.
Our study is focused on textual DSLs, specifically those developed using the Xtext framework \cite{bettini2016implementing}, which is  particularly widely used in the MDE community, due to its rooting in the Eclipse ecosystem.
Xtext also serves as a blueprint for increasingly widely used DSL workbenches such as \textit{textX} (\url{https://pypi.org/project/textX/})  and   \textit{langium} (\url{https://langium.org/}). 

As part of our contribution, we provide a dataset \cite{dataset} of 1002  repositories (via their URLs) together with our extracted meta-data, e.g., the repository's type, employed development scenario,  availability of various artifacts, and change statistics.
This dataset addresses the need for large and consistently documented artifacts expressed in the MDE community \cite{robles2023reflection,damasceno2021quality} and can be particularly useful for follow-up research, both to develop advanced (e.g., AI-based) techniques, as well as supporting the identification of  cases that can be used to inform design and evaluation activities.

Specifically, we focus on the following research questions:


\noindent{}\textbf{\reviseX{RQ1}:} \textit{Are there GitHub projects that use Xtext? Which are these projects?}
We set out to investigate basic information on Xtext-related activities in GitHub repositories, \reviseX{including the type and application domains of these repositories and their contained languages. This research question is broken down into two further research questions.}

\reviseX{\textbf{RQ1.1}: \textit{How to categorize these repositories?} 
This classification will allow us to know which repositories contain well-documented languages, which repositories have developed infrastructure for Xtext-based DSLs, etc. This will be particularly useful for follow-up research (e.g., our RQ2 to 4).}

\reviseX{\textbf{RQ1.2}: \textit{Which trends define the application domains of DSLs in recent years?}
Answering this question can reveal in which domains Xtext-based DSLs are mainly developed, and in which domains DSL development is increasing, and whether it is increasing all the time. For example, data management and databases are increasingly important domains for Xtext-based DSLs. The frequency of language development is related to the evolution of the language. Answering this question can prepare us for studying the evolution of Xtext-based DSLs.
}

\noindent{}\reviseX{\textbf{RQ2}: \textit{What language artifacts for Xtext-based DSLs do these repositories contain?}}

\reviseX{\textbf{RQ2.1}: \textit{What main language  definition artifacts do these repositories contain?} In answering this question, we investigate three key language artifacts of Xtext-based DSLs, namely, grammars, meta-models, and workflow definitions. This allows us to overview of the contained language artifacts in these repositories that are related to Xtext activities, and the answer to this question is particularly useful for follow-up research (e.g., our RQ3 and 4).}

\reviseX{\textbf{RQ2.2}: \textit{Do these repositories contain both grammars and instances that adhere to it?} The answer to this question can reveal which repositories contain both grammar and example instances and whether they contain both. Providing example instances is beneficial for using the DSL, so it is interesting to know whether the repositories already provide example instances that adhere to the grammar.}

\reviseX{\textbf{RQ2.3} \textit{To what extent is Xtext grammar used by example instances?}
When answering this question, we will know how many grammar rules in the grammar are used by the instance files in the same repository, i.e., the coverage of the instance to the grammar. 
The higher the coverage of the grammar by the example instances, the more support people will get when learning and using the DSL.
}

\excl{ \noindent{}\textbf{RQ_instance:} Do those GitHub projects that use Xtext include both the grammar and instances that adhere to it? }

\noindent{}\textbf{\reviseX{RQ3}:} \textit{Which development scenarios for Xtext-based languages are applied in these projects? }
Xtext supports multiple development scenarios that differ in their complexity (discussed later).
There is a question on whether complex development scenarios such as meta-model-driven development are used in practice and thus, need to be supported with dedicated approaches.
Moreover, in the course of answering this question, we discovered a trend of what we call \textit{retrofitting}---creating an Xtext grammar that fits an existing language.

\noindent{}\textbf{RQ4:} \textit{How do Xtext-based languages on  GitHub evolve over time?}
Since DSLs are often envisioned as "small" languages \cite{deursen1998little}, it is tempting to view their evolution as a non-issue.
In this RQ, we study longitudinal aspects of language projects, including their longevity  and amount of changes performed.
We highlight the existence of long-living language projects  and shed light on their proneness to significant changes, leading to challenges that we discuss later, in Sect.~\ref{sec:discussion} of the paper.

\reviseX{\textbf{RQ4.1}: \textit{How can evolution in these projects be characterized quantitatively?}The answer to this question depicts the active time span of the grammar in these repositories, i.e., the time of first creation and the end time of the last evolution (with the commit time as reference). In addition, the answer to this question also reveals the quantity of grammar evolution in these repositories, i.e., the number of changes in grammar rules and the number of changed lines of grammar definitions. These quantitative results depict the evolution of the language in these repositories.}

\reviseX{\textbf{RQ4.2} \textit{How common are different types of changes during the evolution of grammars?}
An important aspect of evolution is the reason for the evolution changes. Software systems change, because of changing requirements and user needs. However, they also change for maintenance reasons, e.g. when adaptations are necessary due to changes in used libraries or operating systems or when failures occur that need to be corrected. We ask the question whether evolution of DSLs is mostly due to maintenance reasons or whether changes to extend the languages are common.   
}

\reviseX{\textbf{RQ4.3} \textit{How do textual instances co-evolve with grammar in real projects?}
Xtext-based DSLs are developed in repositories. Like general-purpose languages, DSLs may also evolve. For those repositories that have developed DSLs, by answering this question, we understand how their instances co-evolve with the grammar. This will provide insights into whether instances are likely to stay up to date with the evolving grammar. 
}





\reviseX{This paper is a significantly extended version of our previous conference paper \cite{zhang2024}, in which we first presented our dataset and addressed a subset the research questions stated above.
For the present manuscript, we considerably extended our analysis to provide a more in-depth look into the included projects, their included artifacts, and evolution activities.
This entailed extensive work on classifying domains and analyzing their trends  (leading to the new RQ1.2), retrieving  language instances and analyzing their coverage of DSL concepts (RQ2.2 and RQ2.3, both new), and manual labeling of commits and studying files contained in them (RQ4.2 and RQ4.3, both new).
Based on the new findings, we also significantly extended our discussion, to discuss the requirements for better supporting artifact co-evolution (significantly extended Sect.~\ref{sec:discuss-coevol}) and the need for a  better understanding of DSL development in practice (new Sect.~~\ref{sec:discuss-underst})
}

\reviseX{The rest of this paper is structured as follows.
In Sect.~\ref{sec:bg}, we introduce  necessary background
In Sect.~\ref{sec:rw}, we discuss related work.
In Sect.~\ref{sec:method} and \ref{sec:results}, we present the  methodology and results for our study.
In Sect.~\ref{sec:discussion}, we discuss implications of our results as well as threats to validity.
In Sect.~\ref{sec:conclusion}, we conclude.}

\section{Background}
\label{sec:bg}

\subsection{Xtext}
The development of DSLs is supported by a large variety of existing workbenches \cite{erdweg2015evaluating}.
In this paper, we focus on the  Xtext workbench \cite{bettini2016implementing},  because we are interested in language development in MDE contexts, which provides various benefits for language development.
First, the possibility to apply a rich ecosystem of existing MDE tools and techniques to the developed languages and their models.
Second,  support for blended modeling \cite{david2023blended},  a rising paradigm in which several different concrete syntaxes (e.g., graphical and textual) are provided for the same underlying meta-model, which allows the developers to choose one that fits best for the task at hand.
In an MDE context, Xtext is the most widely used technology for textual DSL development, and naturally supports blended modeling, since it involves the use of meta-models for abstract syntax specification, and goes beyond meta-modeling by also allowing to define a textual concrete syntax.

Xtext allows the specification of a textual DSL in terms of an extended EBNF grammar, where the extensions are mappings of language elements to an underlying meta-model (in EMF~\cite{steinberg2008emf}).
The metamodel specifies the language's abstract syntax (language concepts and their relations), whereas the grammar specifies the concrete syntax (keywords, parentheses, nesting of elements) with the mapping to the abstract syntax.
From the provided specification, Xtext can automatically generate comprehensive tool support, including a textual editor with automated checks, syntax highlighting, and auto-formatting.
While Xtext is rooted in the Eclipse ecosystem, adapters for other IDEs (e.g., IntelliJ) are available.

In our repository mining context, we identify grammars and meta-models, as well as two additional artifact types,  as distinct file types. \excl{, illustrated in Figure~\ref{fig:xtext-artifacts-types}.}
In Xtext-based DSL development, there are the following four main MDE artifacts: 1) Xtext files\excl{, whose extension is ``xtext''}, 2) Ecore metamodel files\excl{, whose extension is ``ecore''}, 3) modeling workflow engine (MWE) files\excl{, whose extension is ``mwe2'',}, and 4) textual instance files.
MWE files support the orchestration of  automated activities, in particular, the generation of modeling components. Among others, they define the file extension for textual instances (a.k.a. models created using the language), which render them interesting for our study of artifacts. Xtext files \reviseX{(i.e., grammar file)} can be generated from Ecore files, and conversely, Ecore files can also be generated from Xtext files~\cite{bettini2016implementing}. An MWE file is used to generate Xtext artifacts from Xtext grammar~\cite{bettini2016implementing}, and these Xtext artifacts are used to generate a textual editor for the DSL. Textual instance files are edited in this textual editor.

\excl{\begin{figure}[t]
\centering
    \includegraphics[scale=0.7]{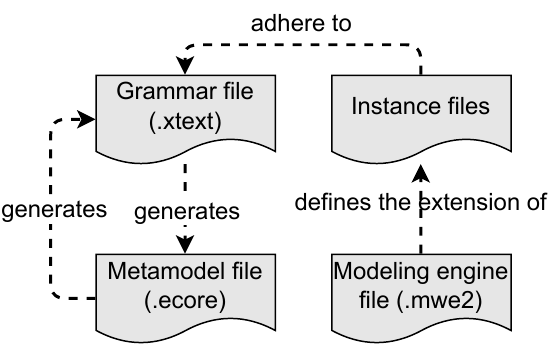}
\caption{Main MDE artifact types in Xtext-based projects.}
\label{fig:xtext-artifacts-types}
\end{figure}}

\subsection{Development Workflows and Scenarios}
\label{sec:scenario}
To specify a DSL's concrete and abstract syntax, Xtext uses two separate artifacts--a grammar and meta-models---, which leads to the challenge of keeping them synchronized with each other as the language evolves.
To this end, Xtext supports two main development workflows \cite{bettini2016implementing}:
In a \textit{grammar-driven} scenario, the user primarily edits the grammar, and has changes propagated  to the meta-model by completely re-generating it.
In a \textit{metamodel-driven} scenario, the user primarily edits the meta-model.
After the changes to the meta-model, the grammar has to be updated, which in the default process has to be done manually--re-generating the grammar is not feasible due to potential information loss about the concrete syntax.

Choosing an appropriate workflow requires to consider the context in which the language is developed.
The meta-model driven workflow is useful in scenarios where the meta-model has to be manually managed, e.g., when it is the center of an already-existing ecosystem of tools, when it comes from a third party vendor or standardization committee, or in a blended modeling scenario with several concrete syntaxes.
The grammar-driven workflow is simpler to support and, therefore, generally preferable in other scenarios.

\subsection{Software Maintenance Intentions}
\label{sec:sw_evo_purpose}
\reviseX{Following a categorization of Lientz and Swanson \cite{lientz1980software}\cite{bennett2000software} the intentions behind software maintenance and evolution can be classified into 4 groups: 
}

\reviseX{
\emph{Adaptive} changes or maintenance describes changes that happened in reaction to a change in the software environment. For example, a change in a used API might require an adaptation of a software system to ensure that it continues working correctly.
}

\reviseX{
\emph{Perfective} changes or maintenance describes enhancements of a software system or reactions to changing requirements, e.g., through changing or adding features and functionality.
}

\reviseX{
\emph{Corrective} changes or maintenance describes corrections of failures and errors of the system, e.g., after a bug report was filed or serious performance issues were discovered.
}

\reviseX{
\emph{Preventive} changes or maintenance describes changes to the software that are meant to ease future changes, e.g., by refactoring the system to ease future adaptations, and help prevent problems before they occur, e.g., by removing code clones to prevent inconsistent changes to the system. }

\section{Related Work}
\label{sec:rw}
\excl{We discuss related work in three directions: on previous language evolution experiences specifically focusing on DSLs, support for co-evolution, and on repository mining for model-driven engineering artifacts.}

\subsection{DSL Evolution}
\label{rw_dslevolution}
The evolution of DSLs has been studied so far with a focus on providing improved evolution support, and on reporting individual cases of evolving DSLs. 
Both are studied in a systematic mapping study of Thanhofer-Pilisch et al.~\cite{thanhofer2017systematic}, the results of which can help researchers and practitioners working on DSL-based approaches obtain an overview of existing research on and open challenges of DSL evolution.
We now discuss selected cases with a particular focus on industrial experiences.
Mengerink et al. investigated the evolution of DSLs in a large industrial MDSE ecosystem~\cite{mengerink2018exploring}, through which they summarized common evolution types and evaluated the automation capabilities of evolution in real-life scenarios. Schuts et al.~reported in~\cite{schuts2021industrial} their experience in evolving a Philips-owned DSL, with the goal of enabling the DSL to support a range of Philips systems. This is concrete experience from the industry regarding the evolution of DSL.
\excl{
Babkin and Ulitin demonstrate the practical use of the proposed projection approach for developing and modifying DSLs in a dynamic context by developing algorithms for a DSL evolution process across model transformations during the development of two case systems~\cite{babkin2023practical}. 

\subsection{Supporting co-evolution in MDE contexts}}
DSL evolution generally leads to issues with keeping multiple involved artifacts synchronized with each other \cite{lammel2018software}.
In the MDE community, a plethora of work  on  co-evolution  problems has evolved, in particular metamodel-model co-evolution \cite{wachsmuth2007metamodel,herrmannsdoerfer2009cope,hebig2016approaches}\excl{
and metamodel-transformation co-evolution
\cite{garcia2012model,khelladi2017exploratory,kusel2015consistent}}, where changes to the metamodel make evolution of the associated models and transformations necessary.
\excl{Recent trends are to offer improved support for these tasks through  AI and user assistants \cite{kessentini2020interactive,kessentini2022semi,di2023modeling}.
A related line of work that addresses the interplay and synchronization of a multitude of models is model federation \cite{guychard2013conceptual,golra2016addressing,drouot2019model}; yet, an application to our considered scope of textual language development and evolution is not available yet.
Co-evolution of artifacts in a textual language context has received comparatively little attention.
The only dedicated work we are aware of is that of Zhang et al.~\cite{zhang2023automated,zhang4379232supporting}, who provide support to automate parts of grammar synchronization in metamodel-grammar co-evolution scenarios.}
As we discuss later, our dataset and results could inform the development of new approaches in this area.


\subsection{Mining}
Due  data availability and volume, GitHub has become the data source of choice for repository mining research~\cite{gousios2017mining}, including those in model-driven engineering. In~\cite{shrestha2023evosl}, Shrestha et al. mined MATLAB/Simulink-related repositories and assembled  a large corpus of Simulink projects, which includes model and project changes and allows redistribution. Mengerink et al. used software repository mining to create a large corpus of OCL constraints~\cite{mengerink2019empowering}. Previous studies of EMF metamodels focused on collecting EMF models from Eclipse projects \cite{kogel2018dataset}, studying the use of meta-modeling concepts in GitHub projects~\cite{babur2024language}, and on deriving a high-quality dataset for machine learning \cite{lopez2022modelset}.
Hebig et al.~\cite{Hebig2016mining} investigated the use of UML in OSS projects by systematically mining GitHub projects.
No previous study focused on Xtext-based DSLs in GitHub repositories. 

The only previous work that explicitly applied repository mining to Xtext grammars (among other MDE artifacts) is MAR, a search engine for models \cite{lopez2022efficient}.
MAR offers a by-example query mechanism for searching a database of 600K models retrieved from existing repositories.
While their underlying dataset includes Xtext models from GitHub, it is not annotated with the metadata offered in our dataset (e.g.,  number of instances, development scenario, evolution statistics).
Moreover, our research contribution and questions have a different scope, focused on characterizing Xtext-specific projects, with their development and evolution scenarios.

\section{Methodology}
\label{sec:method}
We now describe our used repository mining methodology ~\cite{gousios2017mining}.
Like the previous work discussed above, 
our study was focused on  GitHub, the largest existing software repository platform.
Our overall process\reviseX{,  shown in Figure~\ref{fig:overall-process},} is divided into \reviseX{7} steps. First, we obtained a list of non-fork open-source repositories containing Xtext files from GitHub. Second, we cloned all the retrieved open-source repositories to a local hard drive to facilitate access and information acquisition. Third, \reviseX{we manually classified all obtained repositories, before analyzing the collected data with respect to our research questions. Fourth, we searched for relevant file types in these repositories by file extensions and collected information about them. 
Fifth, 
using the repository-specific file extensions of the instances that we identified from MWE files, we collected the instances of these extensions and calculated information about them.
Sixth, 
we analyze the development scenario of each repository.  
Seventh, we analyze the evolution of the languages.}

\reviseX{
\begin{figure*}[htbp]
\centering
    \includegraphics[width=\linewidth]{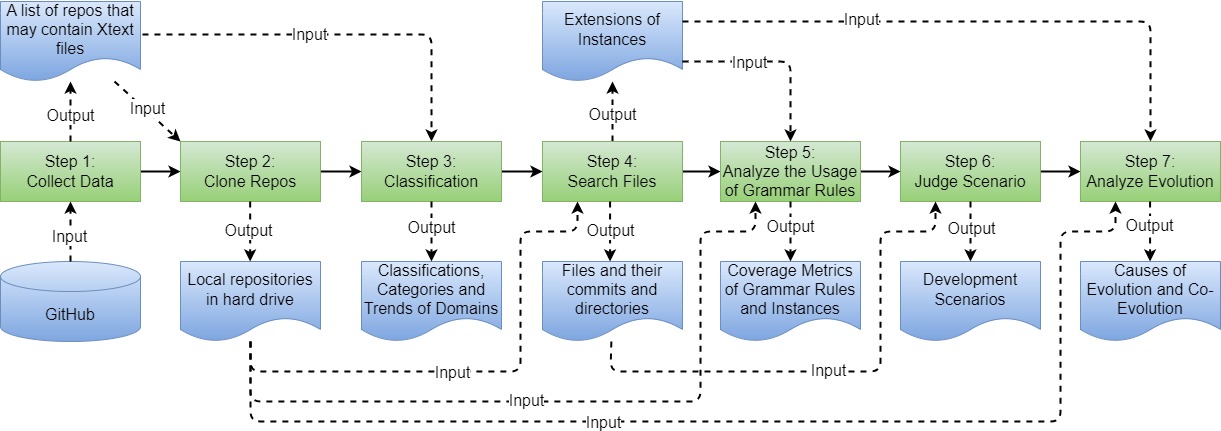}
\caption{Overall process}
\label{fig:overall-process}
\end{figure*}
}

\subsection{Step 1: Data Collection}
We used the GitHub API to obtain repositories that are related to Xtext.
Since the most fundamental MDE artifact of an Xtext project is its grammar, which is stored as a file with the extension \textit{.xtext}, our search string was based on the main clause \texttt{q?=.xtext} -- that is, we searched repositories that contain a file with that extension. We later observed that this search string partially led to the identification of repositories that did not actually contain an Xtext file, but were still related to Xtext in a different way, as described below.
Furthermore,  to exclude repositories that are forks of other repositories, as these might mostly replicate the information from the original repositories and thus bias our results, we set the parameter ``fork" to ``false". 

One complication was that the GitHub API only allows access up to 1,000 results, even when using the \textit{pagination} feature. However,  from a trial search on the GitHub website, we observed that the number of relevant repositories may exceed 1,000.
Hence, added a third parameter to the request, which is the creation time of the repository. We use January 1, 2018, as the boundary to divide the request into two, i.e., requesting results before that date and from that date. 
We retrieved six pages with 576 repositories created before January 1, 2018, and five pages with 426 repositories created from this date, leading to a total of 1002 repositories.

We developed a Python script to complete the above work. Its functions included setting request parameters, sending requests, and dumping request results into local text files. Execution of this script only took a few seconds to complete.
The rationale for developing a new script, instead of starting from an existing dataset (e.g., \cite{lopez2022efficient})  was that it allowed us to retrieve the most up-to-date information from GitHub and that it naturally integrated with our remaining analysis activities.
The overall query used for the requests had the following form:

\begin{quote}
\small
\sloppy
\texttt{https://api.github.com/search/repositories?
        q=.xtext+created:\{since\_date\}..\{stop\_date\}
        +fork:false\&page=\{page\}\&per\_page=100}
\end{quote}

The repository information we obtained contains various information about the repository, such as the repository's ID, name, whether it is private, owner, html\_url, and description. We developed another script to extract the name, owner's login, and html\_url of these repositories from the text files and store them in a table 
to facilitate subsequent data mining and analysis. The result was a table containing 1002 rows.

\subsection{Step 2: Repository Cloning}
GitHub allows to obtain information about GitHub repositories through the GitHub API. However, GitHub has restrictions on the rate and frequency of access. Since we in subsequent analysis needed to frequently access different and large numbers of files,  we decided to clone all repositories to a local hard drive and further analyze the local clones.
We only cloned the \texttt{master} branch (except for a few cases where  \texttt{master}  was empty, where we manually identified a different main branch instead), leaving an analysis of use of branching and pull request as future work.

\excl{To avoid name collisions during cloning arising from several repositories having the same name (e.g., 35 repositories had the name \texttt{xtext}), we renamed each repository after cloning to include the  repository owner name, which leads to a unique name.}


\subsection{\reviseX{Step 3: Classification}}
\label{sec:classification}
\reviseX{In the third step we classified the repositories and domains of the DSLs.}

\reviseX{\subsubsection{Classification of Repositories}}
To give deeper insights into the different kinds of Xtext-related repositories on GitHub, in Step \reviseX{3}, we manually classified repositories into different types and analyzed the frequency of different types.
Our process for this was as follows:
First, one author manually labeled a sample of 200 repositories with improvised labels. The labels emerged from the observations that a few specialized categories were recurring between the repositories, with proper language projects (described below) being of most interest for our study.
Second, in a discussion between the authors,  the obtained labels were harmonized, by defining descriptions and explicit criteria for them.
Third, we labeled the complete set of all repositories with the final set of labels.
The final labeling was done by one author and checked by another author. 
Disagreements were resolved together.

The obtained list of types together with their descriptions and criteria was as follows.\\
\noindent{}\textit{Language}: A repository with proper, documented Xtext-based language. \textit{Criterion:} The README.md or ``About'' section describes it as (implementation of) a language, or a software system that incorporates a clearly identifiable language. \textit{Notes}: We also collected the language's domain. After noticing that several repositories re-implemented an existing language  (a phenomenon we call \textit{retrofitting}), we also noted whether this is a case for each repository. \\
\noindent{}\textit{Training/Examples}: A repository serving the training of Xtext users, usually in the form of an example, tutorial or both. \textit{Criterion:}  The project's README.md or ``About'' section describes it as an example, a tutorial, or a demonstration. \textit{Note}: using the word "example" as part of the name was not deemed as a useful criterion, as examples might be created for experimental purposes---see below. \\
\noindent{}\textit{Infrastructure}: A repository with tooling for supporting development with Xtext.  \textit{Criterion:}  The README.md or ``About'' section suggests that the project is about supporting tooling. \\
\noindent{}\textit{Experimental/Personal}: A repository that does not fall in any of the above categories, but is still directly related to the language workbench Xtext.  \textit{Criteria:} Any of the following applies: 1. The contained grammar is extremely small and basic. 2. The contained grammar is taken from a standard example provided with Xtext. 3. The README.md and ``About'' section are empty or give no context information on what the repository is about. 4. The README.md and ``About'' section describes it as an assignment submission for a course, or as an example for debugging purposes. \\
\noindent{}\textit{Unrelated}:  A repository unrelated to the language workbench Xtext, except for naming. \textit{Criterion:}  The only connection to the language workbench is sharing (parts of) the name.

\reviseX{\subsubsection{Domain Identification and Categories}
\label{sec:category}
}
\reviseX{In RQ1.2, we focused on 226 repositories classified as ``Language''. We manually reviewed these 226 repositories to identify their domains. To reduce the bias caused by manual identification, one of us performed the initial identification for each repository, while one of the other authors reviewed each identified domain.
During the review process, when there was a disagreement, we discussed and voted on the final domain determination within the author team.}

\reviseX{
The resulting list of domains was at a fine-grained level, with entries such as \textit{Security protocol analysis}, \textit{SAT solving and verification}, and \textit{Telemedicine System}, 
To overview the variety of domains and to analyze the trends in application domains over time, we needed to categorize these fine-grained domains into coarse-grained categories, such as \textit{Security and Networking} and \textit{Artificial Intelligence and Machine Learning}.
To automatically derive suggestions for such categories, we adopted an LLM-based method proposed by Chen et al.~\cite{chen2023prompting}. In our case, the used LLM was ChatGPT-4.0. We fed our list of 226 fine-grained domains into a prompt of the form ``\textit{I have 226 domains, which are placed in the attached TXT file. Could you categorize these 226 domains into larger categories?''} We repeated this approach five times, manually reviewed the resulting lists and voted on them to determine the best suggestion. This resulted in the following list of categories: ``Programming Languages'', ``Software Development and Engineering'', ``Games'', ``Web and Mobile Development'', ``Modeling, Simulation, and Design'', ``Data Management and Databases'', ``Security and Networking'', ``Artificial Intelligence and Machine Learning'', ``Business and Enterprise Applications'', ``Healthcare and Life Sciences'', ``IoT, Embedded Systems, and Hardware'', ``Mathematics, Logic, and Scientific Computing'', ``Testing and Verification'', ``Content, Information and Document Management'', ``Cloud, APIs, and Web Services'', ``Graphical User Interfaces (GUI)'', ``Questionnaire'', and ``Miscellaneous''.}

\reviseX{We then manually mapped the 226 domains to these 18 categories. This was again done first by one of us and followed up by a review of the mapping performed by another author. 
Again, if there was a disagreement, we decided on the final mapping by discussing and voting within the author team.}

\reviseX{}

\subsection{Step 4: File Search}
\label{sec:file_search}
\reviseX{\subsubsection{Xtext/Ecore/MWE2 File Search}}
Given the local clones of the 1002 identified repositories, in Step \reviseX{4}, we  
identified their contained grammars, meta-models, and workflow files, by searching for files with the extensions ``.xtext'', ``.ecore'', ``.mwe2'', respectively.

\excl{We performed the same retrieval and computation on metamodel files with the ``.ecore'' extension and modeling workflow engine files with the ``mwe2'' extension.} 
\reviseX{ Furthermore, we did additional analysis, specifically, whenever we found an Ecore metamodel file, in addition to recording the number of commits, we also recorded the name of the folder containing the file for subsequent analysis of the language development scenario; whenever we found an MWE2 file, in addition to recording the number of commits, and we also looked for the instance extension defined in it.}

\excl{In an MWE file, the extension of a textual instance is usually defined in the field of \texttt{fileExtensions}. However, it is also common practice to define the extension of the instance through the global variable \texttt{file.Extensions} before the \textit{workflow} code block. Textual instance extensions can also be defined outside the MWE file, or they can also be defined with non-English letters, we exclude both of these two cases.}

\reviseX{\subsubsection{Instance Search}}
One Xtext artifact type we set out to study was instances (models); yet, identifying instances is non-trivial, as their file extension differs per language.
We identified instances by reading the file extension from the previously identified MWE files and then searching relevant files.
To check whether the found instances adhere to the grammar in the same repository, we performed a sampling analysis: we randomly selected ten repositories that contained both grammar and instances and manually checked conformance. \reviseX{For all ten repositories we found that the contained instances fully conformed to the contained grammars. Thus we will make the assumption that instances files found in a repository with an Xtext grammar are most likely written in the language specified by that grammar.}



\reviseX{\subsection{Step 5: Analyze the Usage of Grammar Rules}}
\label{sec:used_grammar_rule}
To answer RQ2.3, we need to obtain the total number of grammar rules in each grammar (i.e., the xtext file) and the number of types of objects in the instance (i.e., the grammar rules used). To this end, \reviseX{for those repositories categorized as ``Languages'', we first filter out those that contain both grammars and instances. We} developed a script that counts the total number of grammar rules in all Xtext files in a single repository. When counting grammar rules in xtext files, we skipped the xtext files in the paths \texttt{src-gen} and \texttt{bin} because they are backups of the xtext files in \texttt{src}.

At the same time, \reviseX{for each grammar, we need to obtain each element in the instance that complies with it and the type of these elements. The types in the instance correspond one-to-one to the grammar rules in the grammar. Obtaining the total count of all types used in the instances and the total count of grammar rules in the grammar can be used to calculate the coverage of the instance to the grammar. Our script}
can traverse all child elements in the found instance and obtain their types. Then place the types in a list by removing the duplicate ones. Ensuring that the project where the Xtext grammar is located has no errors and can be resolved normally is a prerequisite for obtaining the element types in the instance. A technical problem encountered was that the Xtext projects in the repository could not be resolved on the experimental machine (i.e., the author's computer) due to Xtext version issues. The Xtext version on the experimental machine is 2.36, while many repositories contain the Xtext projects that were created with an older version of Xtext. Given the practical difficulties of downloading and installing old versions of Xtext, we re-created these Xtext projects with Xtext 2.36 by fully reusing their grammars. Then we used our script to call the re-created Xtext projects to parse the instances in the original repository to calculate the number of used grammar rules.

\reviseX{Finally, for each repository, we divided the sum of the number of grammar rules by the total count of collected non-repeated types across all instances to obtain the percentage of grammar rules used in that repository.}

\reviseX{}

\subsection{Step 6: Scenario Judgement}
In Step 4, we determined the used language development scenarios in each repository. 
Xtext supports two language development scenarios, described in Sect.~\ref{sec:scenario}. In the grammar-driven scenario, the text definition of the Xtext grammar has a statement starting with the keyword ``generate", which results in generating a meta-model from the grammar. When generating the metamodel, Xtext automatically places the metamodel in a folder named ``generated''. In the metamodel-driven scenario, language developers create a metamodel in a folder they create. We assumed that the developers do not name their created folders ``generated'', which would be counterintuitive. 
Considering that there may be multiple Ecore files in a repository, we distinguished three cases in which Ecore files existed in a repository: 1) All Ecore files in the repository are in a folder named ``generated", 2) all the folders containing Ecore files in the repository are not named ``generated", and 3) some of the folders containing Ecore files in the repository are named ``generated" and some are not. We classify the first situation as a grammar-driven scenario, the second situation as a meta-model-driven scenario, and the third situation refers to both scenarios.

\subsection{Step 7: Analyze Evolution}
\label{sec:step_7}
\reviseX{To analyze the evolution and co-evolution in the 226 repositories classified as ``Language'', we analyze the commits of all Xtext/Ecore files and instance files in these repositories. First, we obtained the commits of these files in the 226 repositories using a Python script, including the commit time and message. The commit information was stored in an xlsx table (both the script and the table can be found in our supplemental materials~\cite{dataset}). 
Not every commit includes an evolution step in and of itself. For example, changes to a language (as with all software systems) will often be complex enough to warrant effort over multiple days or even weeks, resulting in multiple commits. To study evolution in this paper we therefore introduce a heuristic allowing us to approximate the occurrence of new evolution steps. For that, we calculated the time difference between each commit and the previous commit of the same file in days.
If a commit of a file happens more than 30 days after the previous commit of the same file, then we consider this commit to be the first commit of a new evolution step for that file. 
If a change to a file is committed within five days of the previous commit for that file, then we consider that commit to be an iteration. Differences between five and 30 days could hint at both, iterations and evolution. 
}

\reviseX{Commits that are the start of a new evolution step were further analyzed with regard to the purpose of the commit.
We divided these commits into adaptive, perfective, corrective, and preventive types based on the intentions of software maintenance and evolution described in Section~\ref{sec:sw_evo_purpose}. To do so we manually analyzed the commit comments of these commits as well as commit comments from iterations, i.e., commits following shortly after the analyzed commit.
We marked commits that do not have sufficient information for us to determine their type as ``unclear''.}

\begin{figure}[tb]
\centering
    \includegraphics[width=\textwidth]{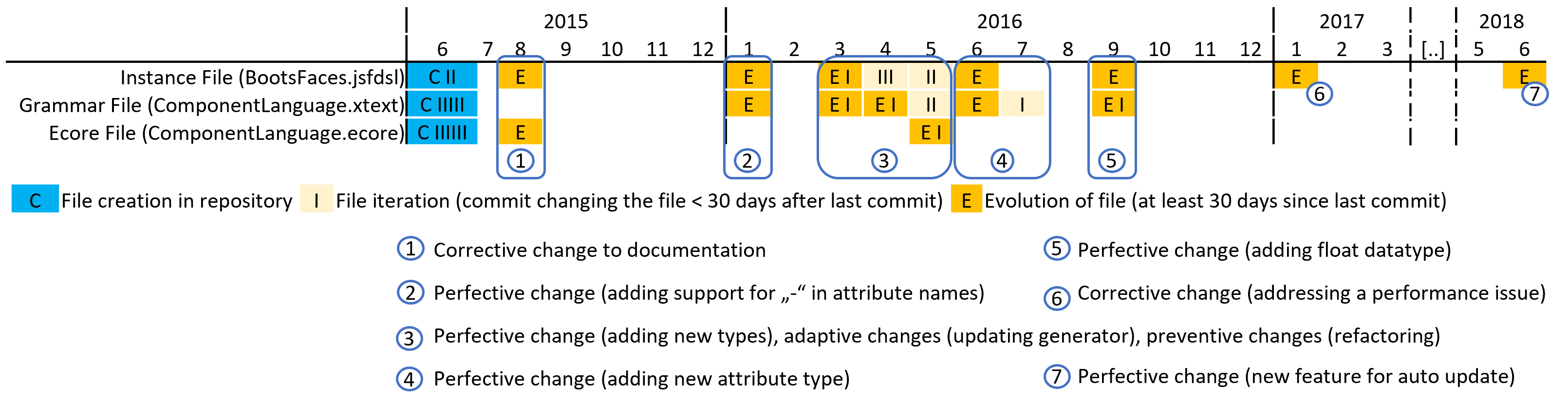}
\caption{Example of Evolution History.}
\label{fig:evolution_history}
\end{figure}

\reviseX{
As an example, Figure~\ref{fig:evolution_history} illustrates the timeline of the creation and changes of three files from the repository JSFLibraryGenerator\footnote{\url{https://github.com/stephanrauh/JSFLibraryGenerator}}: an ecore file, a xtext file and an instance file conforming to the grammar.  
All three files were initially committed in the June of 2015 directly followed by several iterative commits changing the files. In August 2015 an evolution step happened that included a corrective change to the documentation and caused changes in the ecore and instance files. Further evolution changes happened in 2016, January 2017, and June 2018. In the last two cases only the instance file changed. The evolution step in March 2016 starts with a simultaneous change of the grammar and instance file. The instance file is then changed over a series of several commits (iterations). The xtext file is iterated on as well, then a break of more than 30 days occurs and the xtext file is changed again, which is counted as a new evolution step in our heuristic. Interestingly, the iterations of the instance file continued during this 30 day period. This example shows that it can be difficult to be sure that a change after more than 30 days really presents an independent evolution step. There are multiple types of changes occurring during that period as well, including perfective, adaptive, and preventive changes. 
Nonetheless, in most of the cases the heuristic of 30 days allows us to approximate developmental phases, which we consider here to be evolution.
}


\reviseX{As can be seen in Figure~\ref{fig:evolution_history} some evolution steps of a pair of grammar and instance files (and ecore file) occur within days or even within the same commit. 
To better understand the prevalence of co-evolution of grammars and instances, we conducted a separate investigation on commits starting an evolution step (i.e., commits that changed a grammar or instance file that has last been changed at least 30 days earlier).
}
\reviseX{
For each commit, we analyzed whether a file of the other type (instance file or xtext file) in the same repository has been changed within the same commit or within five days.
If so we checked whether the change of that other file presents the start of an evolution step as well (i.e., is not an iteration following other changes).
In our example, there is a co-evolution of grammar and instance file in March of 2016. However, the second evolution step of the xtext file in April of 2016 is here not considered a co-evolution with the instance file, even though it changed in close time proximity, since the instance file underwent a continuous iteration starting March of 2016.
}

\section{Results}
\label{sec:results}
\begin{figure}[tb]
\centering
    \includegraphics[scale=0.44]{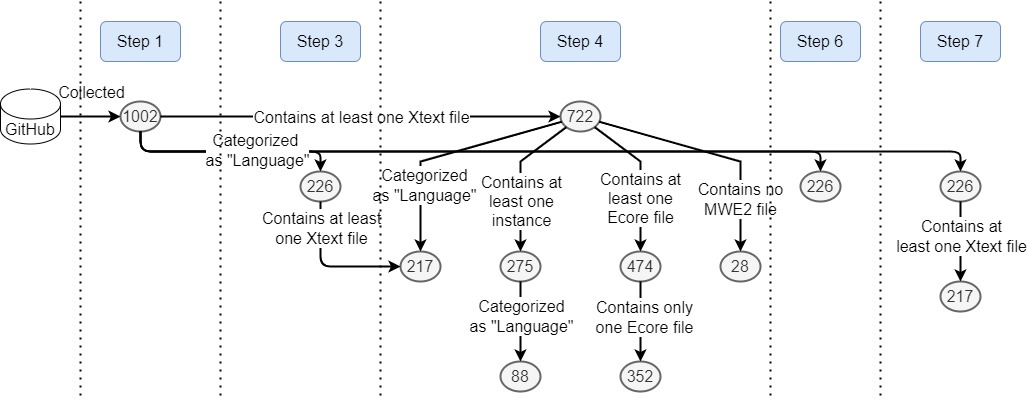}
\caption{\reviseX{Count of repositories in different steps.}}
\label{fig:repos_cnt_diff_steps}
\end{figure}

This section presents the results of our investigation. In this research, an ample amount of data is collected and analyzed, in both manual and automated ways. The automated way used 
scripts developed by the authors. 
The resulting dataset (spreadsheet) 
and our analysis scripts are available from the associated artifact  \cite{dataset}. 
In this section, we will introduce the counts of repositories we obtained by filtering under different conditions in different steps. There are multiple such counts, and we depict the relationship between different counts in Figure~\ref{fig:repos_cnt_diff_steps} for easy understanding.

\subsection{\reviseX{RQ1: Are there GitHub projects that use Xtext? Which are these projects?}}

\subsubsection{\reviseX{RQ1.1: How to categorize these repositories?}}

\begin{figure}[tb]
\centering
    \includegraphics[scale=0.45]{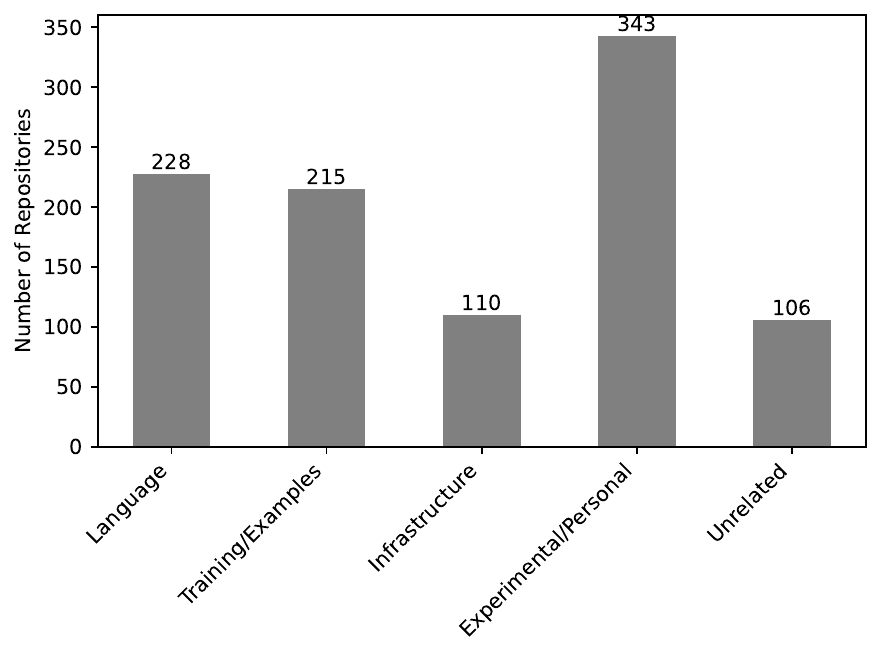}
\caption{Classification of repositories.}
\label{fig:repo-types}
\end{figure}

\revised{To overview the 1002 repositories identified via our search methodology, we show the outcome of our manual classification, according to the methodology explained in Section~\ref{sec:classification}.
As indicated in Figure~\ref{fig:repo-types},  we found  226 repositories in which languages have been developed and contain descriptions of them. There are 215 repositories with documented training and example materials.
110 repositories provide infrastructure to support development with Xtext.
Repositories classified as \experimental are the most numerous, with 343 cases.  Additionally, 106 repositories have no relationship to Xtext except for naming.}

We illustrate the different categories with examples.
\textit{Languages} developed with Xtext span a large variety of cases, from DSLs for specific application domains such as games (\texttt{Casino}), telemedicine (\texttt{telemed}), document management (\texttt{Xarchive}), 
to technical domains such as JSON schema (\texttt{xtext-json}), Quantum computing (\texttt{Quingo/compiler\_xtext}) 
and Eclipse launch configurations (\texttt{lcdsl}).
A noteworthy sub-category, 
are cases of retrofitting existing languages, such as GraphQL (\texttt{graphQL-xtext-grammar}), or 
Oberon (\texttt{Oberon-XText}).
\textit{Training/Example} projects comprise tutorials such as 15-minute Xtext tutorial in Chinese (\texttt{xtext\_tutorial\_15\_min\_zh}).
\textit{Infrastructure} projects include technology for integrating Xtext and particular languages in specific contexts, e.g.,  build processes  (\texttt{gradle-xtext-generator}) and editors (\texttt{vim-xtext}).
\textit{Experimental/personal} code often involves a dump of the user's personal workspace (e.g. \texttt{Xtext Workspace}).
\textit{Unrelated} repositories generally result from a name clash, such as using the name `xtext' for some unrelated tool (e.g.,  \texttt{resloved}'s  text displaying tool \texttt{xtext}), or within some longer name, such as \texttt{TopXTextUI}.


As useful meta-information to inform the identification of particularly widely used repositories, we collected the number of forks and stars for all repositories and included them in our dataset \cite{dataset}.
We found that  218 repositories had forks and  321 repositories had stars.

\reviseX{\subsubsection{RQ1.2: Which trends define the application domains of DSLs in recent years?}}

\begin{figure}[htbp]
\centering
    \includegraphics[scale=0.25]{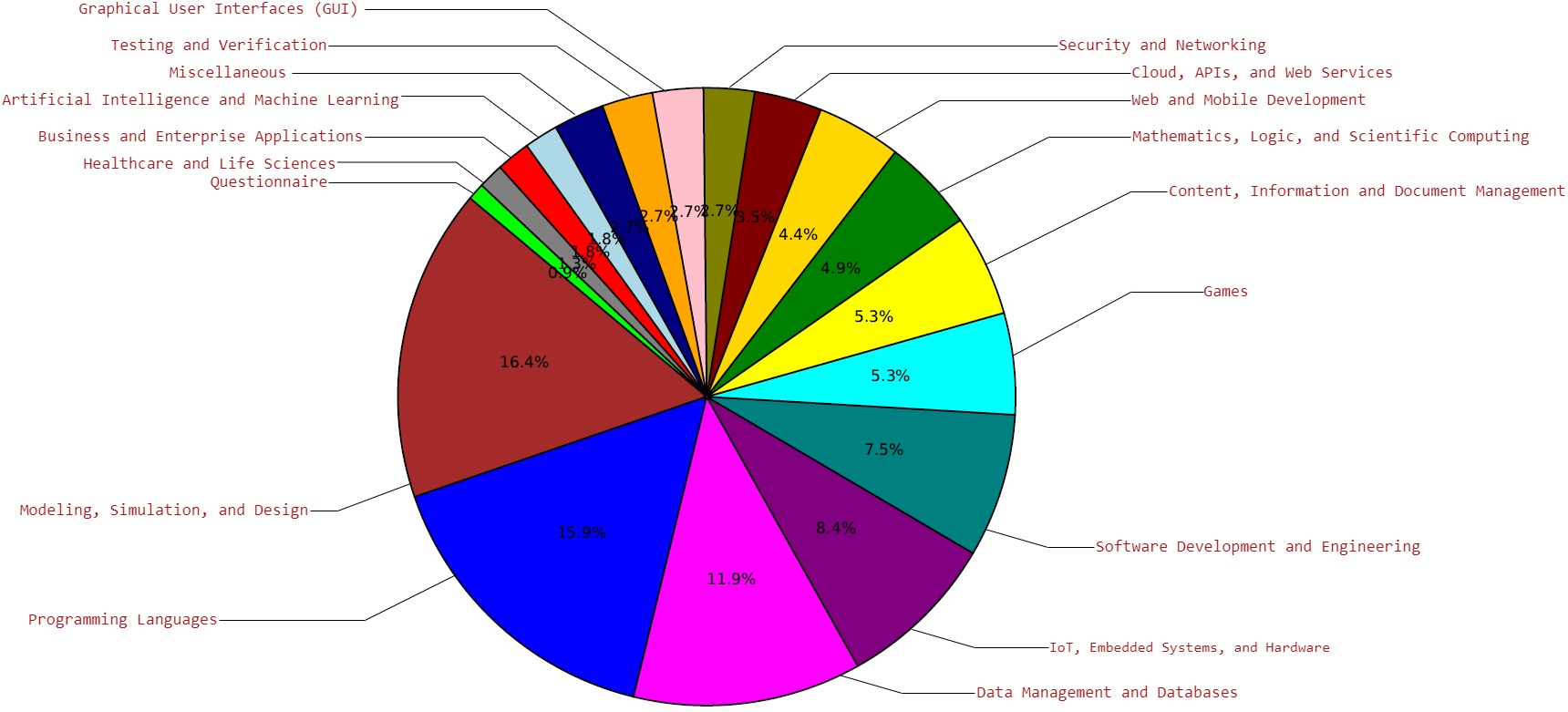}
\caption{Proportion of repositories in different categories.}
\label{fig:repo-diff-cate}
\end{figure}

\reviseX{As mentioned in section~\ref{sec:category}, we mapped the domains of the 226 repositories classified under “Language” to the 18 categories, and counted the number of mapped repositories under each category. The obtained results are shown in Table~\ref{tab:categories}. The distribution of the categories is shown proportionally in Figure~\ref{fig:repo-diff-cate}.
\begin{table}[htbp]
\centering
\caption{Count of mapped repositories under different categories.}
\label{tab:categories}
\begin{tabular}{c|l|l}
\hline
\textbf{No.} & \textbf{Category} & \textbf{Count} \\
\hline
1   & Programming Languages                         & 36 \\
2   & Software Development and Engineering          & 17 \\
3   & Games                                         & 12 \\
4   & Web and Mobile Development                    & 10 \\
5   & Modeling, Simulation, and Design              & 37 \\
6   & Data Management and Databases                 & 27 \\
7   & Security and Networking                       & 6 \\
8   & Artificial Intelligence and Machine Learning  & 4 \\
9   & Business and Enterprise Applications          & 4 \\
10  & Healthcare and Life Sciences                  & 3 \\
11  & IoT, Embedded Systems, and Hardware           & 19 \\
12  & Mathematics, Logic, and Scientific Computing  & 11 \\
13  & Testing and Verification                      & 6 \\
14  & Content, Information and Document Management  & 12 \\
15  & Cloud, APIs, and Web Services                 & 8 \\
16  & Graphical User Interfaces (GUI)               & 6 \\
17  & Questionnaire                                 & 2 \\
18  & Miscellaneous                                 & 6 \\
\hline
\end{tabular}
\end{table}
}

\reviseX{Figure~\ref{fig:repo-diff-cate} shows that half of the Xtext-based DSLs are used in the domains of ``Modeling, Simulation, and Design'' (e.g. language \texttt{Yaktor} in the repository \texttt{yaktor-dsl-xtext} which is for \emph{Data and behaviour modeling}), ``Programming Languages'' (e.g. language \texttt{Quingo} in the repository \texttt{compiler\_xtext} which is a language for \emph{Quantum programming}), ``Data Management and Databases'' (e.g. language \texttt{xtext-orm
} in the repository \texttt{xtext-orm} which is for defining general ORM models), and ``IoT, Embedded Systems, and Hardware'' (e.g. language \texttt{JKind} which is for \emph{Embedded Systems Modeling and Verification}). 
Despite its currently increasing relevance, the domain of AI/ML is so far rarely addressed with DSLs built with Xtext.}

\begin{figure}[htbp]
\centering
    \includegraphics[scale=0.45]{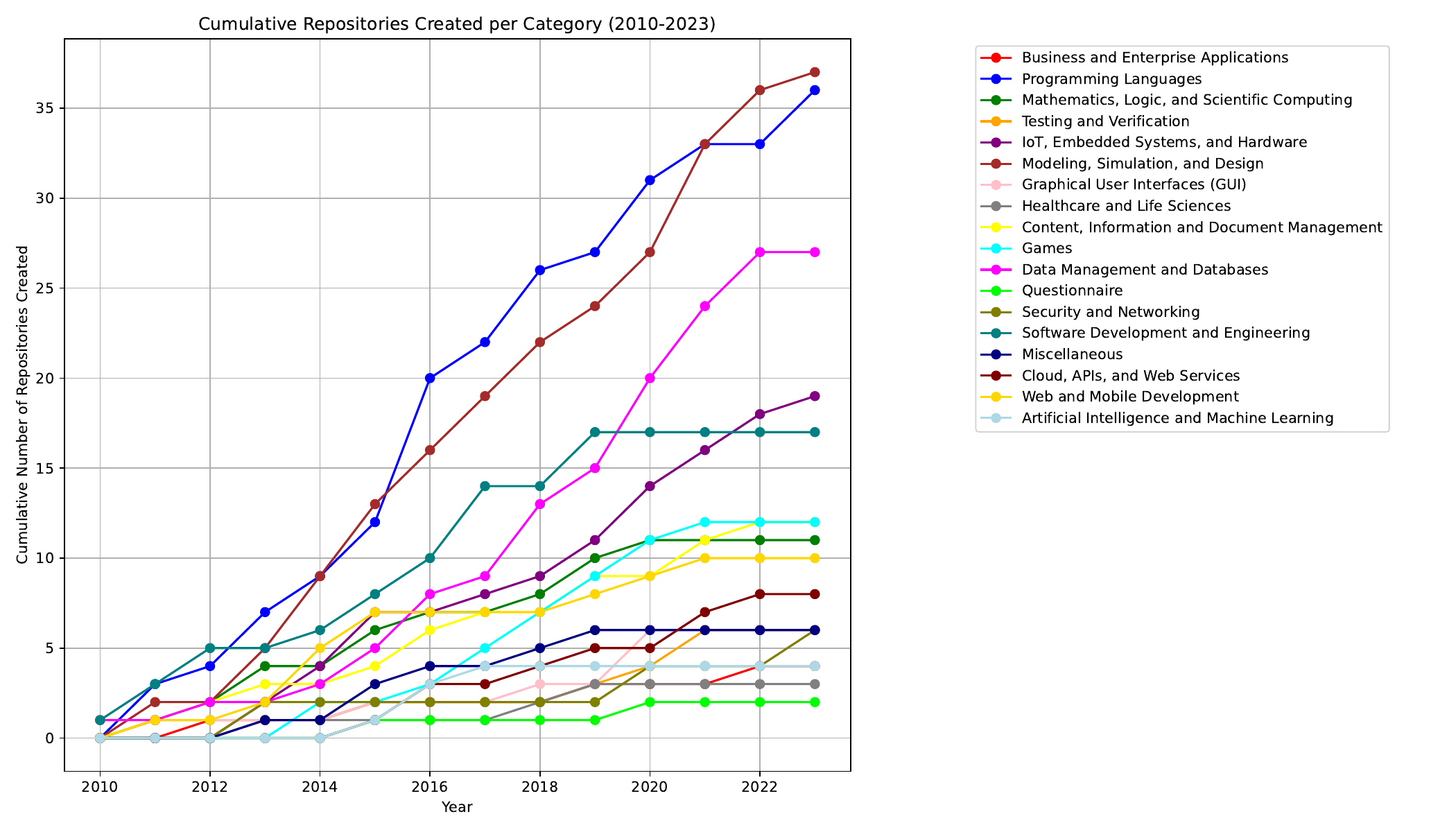}
\caption{Cumulative repository created per category (varies with year).}
\label{fig:repo-per-cate}
\end{figure}

\reviseX{We obtained the creation time of these 226 repositories, and the results showed that the earliest repository was created in 2010, and the last repository was created in 2023. We observed the change trends of these repositories over a span of 14 years from 2010 to 2023, which are shown in Figure~\ref{fig:repo-per-cate}. We found that starting in 2014, Xtext-based language development entered a period of rapid development, and a large number of repositories were created for developing Xtext-based languages. However, after about 2020, Xtext-based language development entered a period of decline, and the number of repositories under most categories stopped increasing. However, the three categories with the largest proportion, i.e., ``Modeling, Simulation, and Design'', ``Programming Languages'', and ``Data Management and Databases'', still maintained an increase in the number of repositories they contained. Interestingly, after about 2019, the development of Xtext-based languages related to ``IoT, Embedded Systems, and Hardware'' became more frequent, while at about the same time, the development of Xtext-based languages related to ``Software Development and Engineering'' suddenly stagnated.}

\reviseX{We also investigated the last commit time of the repositories under these categories. We found that the repositories in nearly half of the categories stopped updating before 2022. The earliest to stop updating were the repositories under the ``AI and ML'' category, whose last commit occurred in 2017. The categories shown in Figure~\ref{fig:repo-per-cate}, where new repositories were still created in recent years, include repositories that are still updated recently. That is, categories such as ``Modeling, Simulation, and Design'', ``IoT, Embedded Systems, and Hardware'', ``Software Development and Engineering'', and ``Data Management and Databases'' contain repositories that still have commits in 2024.}


\begin{tcolorbox}[colback=white!95!black, colframe=white!50!black, arc=3mm, left=0.5em, right=0.5em, top=0.5em, bottom=0.5em]
    \textbf{Results of RQ1: We find five categories of repositories---language, training/examples, infrastructure, experimental/personal and unrelated, and there are 226 repositories classified as ``Language''. From the perspective of domain, these 226 repositories can be categorized into 18 domain classifications. Among them, nearly half of these 226 repositories are categorized into the ``Programming languages'', ``Modeling, simulation, and design'', and ``Data management and databases'' domains, and they are also the three domains with the fastest development and longest active life span of Xtext-based DSLs.}
\end{tcolorbox}

\subsection{\reviseX{RQ2: What language artifacts for Xtext-based DSLs do these repositories contain?}}

\subsubsection{\reviseX{RQ2.1: What main language definition artifacts do these repositories contain?}}

With Xtext grammars being one of the core artifacts of Xtext projects, we checked how many are contained in each repository.
The results,  shown in Table~\ref{tab:file_counts}, show that of these 1002 repositories, only 722 really contain at least one xtext file. \excl{436 of them contain only one xtext file, while 286 of them contain at least two xtext files. Of the 286 repositories, six contain more than 100 xtext files.}
Unsurprisingly, nearly all repositories classified as \lang contained an Xtext grammar---\revised{217 out of 226 cases}.
The nine exceptions involved repositories that contained Xtext-based editors for a particular language without providing the underlying grammar, such as  \texttt{Palladio-Editors-VSCode}, and ports of originally Xtext-based DSLs to other workbenches, such as  \texttt{eJSL-MPS} \cite{priefer2021applying}.


\begin{table}[tb]
    \centering
    \caption{File type frequency in projects containing Xtext files.}
    \label{tab:file_counts}
    \begin{threeparttable}
    \begin{tabular}{c l rl r l r r}
      \toprule					
        ~ 	& \multicolumn{2}{c}{\textbf{Xtext}} & \multicolumn{2}{c}{\textbf{Ecore}} &	\multicolumn{2}{c}{\textbf{MWE}}\\
       \cmidrule(lr){2-3}\cmidrule(lr){4-5}\cmidrule(lr){6-7}
        \#Files  	& \#Repo 	& Perc. & \#Repo & Perc. & \#Repo & Perc.\\
        \midrule
        >= 100 	    & 6     & 0.83\%  & 6     & 0.83\%  & 1     & 0.14\% \\       
        \midrule
        10 - 99 	& 19    & 2.63\%  & 8     & 1.11\%  & 19    & 2.63\%\\ 
        \midrule
        2 - 9 		& 261   & 36.15\% & 108   & 14.96\% & 264   & 36.57\%\\ 
        \midrule
        1           & 436   & 60.39\% & 352   & 48.75\% & 410   & 56.79\% \\ 
        \midrule
        0           & /     & /       & 248   & 34.35\% & 28    & 3.88\% \\
      \bottomrule
    \end{tabular}
    \end{threeparttable}
\end{table}

For the 722 repositories that contain at least one Xtext file,   Table~\ref{tab:file_counts} reports the numbers of other included MDE artifacts. Of these repositories, 248 contain no Ecore metamodel file, while 352 repositories contain one Ecore metamodel file.
A total of 6 repositories contain at least 100 
Ecore files each. All of these are infrastructure projects that use a larger number of examples for testing and/or demonstration purposes; three of them associated with the official Xtext project.
\excl{Pertaining workflow engine files, 28 repositories contain no MWE files, while 410 contain one MWE  file, and one contains more than 100 MWE files.}
\reviseX{We also counted the number of MWE2 files in the repositories, as shown in Table~\ref{tab:mwe2_count}. We found that among the 722 repositories that contained at least one Xtext file, all but 28 contained at least one MWE2 file. The vast majority of repositories contained only one MWE2 file, and only one repository contained more than 100 MWE2 files. We also found 222 repositories that contained MWE2 files but no Ecore files.}
\reviseX{Such a big}
number of projects that contain an MWE file but no Ecore file can be explained by the fact that in a grammar-driven scenario, Ecore files can be fully automatically generated from the underlying grammar, and it is a common practice to not commit automatically generated artifacts to repositories.


\begin{table}[h]
\centering
\caption{Count of MWE2 files in the repositories that contain at least one Xtext file.}
\label{tab:mwe2_count}
\begin{tabular}{c|c|c}
\hline
\textbf{Count of MWE2 files} & \textbf{Count of Repos} & \textbf{Percentage} \\
\hline
>= 100      & 1     & 0.14\% \\
10 - 99     & 19    & 2.63\% \\
2 - 9       & 264   & 36.57\% \\
1           & 410   & 56.79\% \\
0           & 28    & 3.88\%\\
\hline
\end{tabular}
\end{table}


\excl{ \begin{table}[tb]
\centering
\caption{Repositories with the highest number of stars: top 5}
\label{tab:top_five_star}
\begin{tabular}{l|l|r}
\hline
\textbf{Owner} & \textbf{Repo} & \textbf{\#Stars} \\
\hline
eclipse    & xtext  & 746 \\
eclipse      & xtext-core  & 117 \\
eclipse        & xtext-xtend & 101 \\
applause & applause & 98 \\
LorenzoBettini           & packtpub-xtext-book-2nd-examples & 68 \\
\hline
\end{tabular}
\end{table}

\begin{table}[tb]
\centering
\caption{Repositories with the highest number of forks: top 5}
\label{tab:top_five_fork}
\begin{tabular}{l|l|r}
\hline
\textbf{Owner} & \textbf{Repo} & \textbf{\#Forks} \\
\hline
eclipse    & xtext  & 314\\
eclipse      & xtext-core  & 98 \\
eclipse           & xtext-eclipse & 80 \\
eclipse        & xtext-xtend & 53 \\
xtext & seven-languages-xtext &  47 \\
\hline
\end{tabular}
\end{table}}

\subsubsection{\reviseX{RQ2.2: Do these repositories contain both grammars and instances that adhere to it?}}

Table~\ref{tab:instance} shows the results of identifying textual instances in the 722 repositories that contain at least one Xtext file, 447 contain no textual instances at all, 103 contain only one textual instance, and 173 repositories contain at least two textual instances. Interestingly, we found two repositories containing more than 1000 textual instances, namely, \texttt{xtext/xtext-monorepo} and \texttt{eclipse/xtext}, owned by the Xtext team and the Eclipse Foundation, respectively.

\begin{table}[tb]
\centering
\caption{Instances in repositories that contain Xtext grammars.}
\label{tab:instance}
\begin{tabular}{c|l|r}
\hline
\textbf{Count of Instances} & \textbf{\#Repos} & \textbf{Percentage} \\
\hline
>= 1000      & 2   & 0.28\% \\
100 - 999    & 10  & 1.39\% \\
10 - 99      & 30  & 4.29\% \\
2 - 9        & 130 & 18.00\% \\
1            & 103 & 14.27\% \\
0            & 447 & 61.77\% \\
\hline
\end{tabular}
\end{table}

\reviseX{As shown in Table~\ref{tab:instance}, there are 275 repositories that contain at least one Xtext file and at least one instance file. In Section~\ref{sec:file_search}, we mentioned that in order to check whether the found instances comply with the grammar in the same repository, we performed a sampling analysis, that is, we randomly selected ten repositories that contain both grammars and instances and checked the compliance. The results of this sampling analysis show that the text instances in these ten repositories all comply with the grammar in the same repository. Thus, we work with the assumption that the found instances conform to the grammars.}

\reviseX{\subsubsection{RQ2.3: To what extent are Xtext grammars covered by example instances contained in these repositories?}}

\reviseX{Among the repositories categorized as ``languages'', there are 88 repositories that contain both at least one xtext file and at least one instance file. In Step 5, we recreated the Xtext projects contained in some repositories using Xtext 2.36. This resulted in 71 of those 88 repositories where the Xtext projects could be fully resolved and parse instances from the same repository. In the other 17 repositories, we could not recreate the Xtext projects they contained due to technical barriers or missing files, such as other Ecore files that they depended on were not provided. 
In the 71 repositories where we could parse the instances, we used the Xtext projects to parse instances from the same repository to obtain the count of used grammar rules and calculated the percentage, resulting in the results shown in Figure~\ref{fig:repo-used-gr}.}

\reviseX{
\begin{figure}[tb]
\centering
    \includegraphics[scale=0.6]{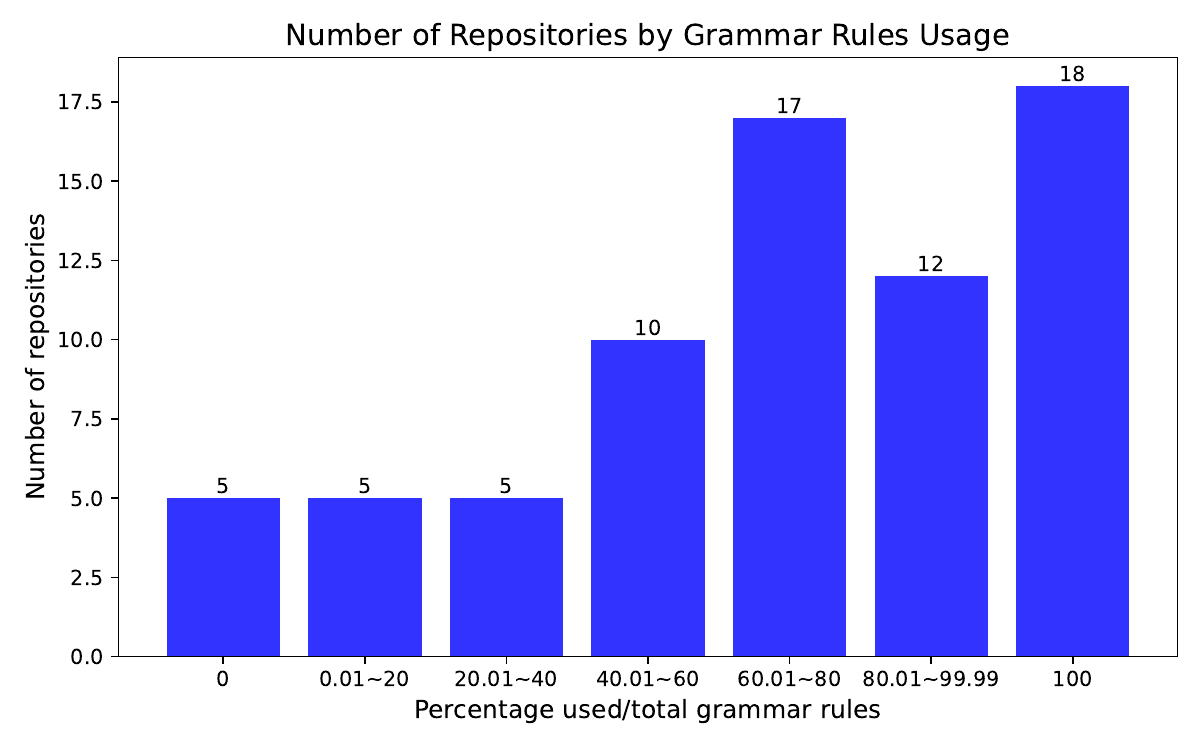}
\caption{Number of Repositories by Grammar Rules Usage.}
\label{fig:repo-used-gr}
\end{figure}
}

\reviseX{We found that among these 71 repositories, most of them provided textual instances that used more than 60\% of the grammar rules. Four repositories provided textual instances, but the textual instances were empty or had nothing to do with the grammar at all. For example, in the repository ``Bicycle-Shop'' with the login value ``ksza'', the extension of the instance file is defined as ``nbs''. The repository does provide a ``nbs'' file, however, the content of the instance does not comply with the grammar.}

\begin{tcolorbox}[colback=white!95!black, colframe=white!50!black, arc=3mm, left=0.5em, right=0.5em, top=0.5em, bottom=0.5em]
    \textbf{Results of RQ2: We found that 722 repositories contain at least one Xtext file. With the exception of a few, almost all of them contain at least one MWE2 file, and the majority of them contain at least one Ecore file. About a third of them contain at least one instance file. We found that most of the repositories that provide instance files use more than 60\% of the grammar rules in the instances.}
\end{tcolorbox}

\subsection{\reviseX{RQ3}: Development Scenarios}
To answer RQ3, on development scenarios for language development, we specifically focused on those repositories classified as \lang and judged their development scenarios as described in Section~\ref{sec:method}, leading to the results shown in Table~\ref{tab:scenario}.

\begin{table}[tb]
\centering
\caption{Frequency of language development scenarios.}
\label{tab:scenario}
\begin{tabular}{c|l|l}
\hline
\textbf{Scenario} & \textbf{\#Repos} & \textbf{\#Retrofitting} \\
\hline
grammar-driven      & 169   & 66 \\
metamodel-driven    & 41    & 6 \\
both                & 9     & 0 \\
not applicable      & 9     & 2 \\
\hline
\end{tabular}
\end{table}

There are a total of \revised{226} repositories classified as \lang, the majority of them (i.e. \revised{169}) are in a grammar-driven scenario, and \revised{66} of them have developed an Xtext version of an existing language. \revised{41} of the 226 repositories are in a metamodel-driven scenario, and \revised{6} of them have developed an Xtext version of an existing language. \revised{Nine} of these \revised{226} repositories contain both grammar-driven and metamodel-driven scenarios, and none of them have developed an Xtext grammar for any existing language. Additionally, the \revised{nine} repositories that were classified \lang but did not contain Xtext files (described in the results for RQ1), were not suitable for our analysis and hence excluded from it.

We give selected examples for the identified metamodel-driven cases since it is arguably the more complex scenario, involving manual overhead for keeping meta-models and grammars synchronized.
\texttt{megal-xtext} provides a textual syntax for the MegaL mega-modeling language, a language that by design provides several concrete syntaxes \cite{favre2012modeling},  a typical motivation for the metamodel-driven scenario.
\texttt{Kotlin-Meta-Model} is a repository with the main claim of providing a meta-model for the Kotlin JVM language;  the provided Xtext grammar is mentioned as an additional artifact.
Other repositories such as \texttt{QuestionnaireDSL} do not include a definite explanation for following the metamodel-driven scenario, but at least contain a visualization for the metamodel (\textit{aird} file), which indicates an intention to explicitly design the metamodel.
Repositories classified as \textit{both} generally comprise several languages with different workflows, e.g., \texttt{telemed} with separate languages for information storage and querying of telemedicine information. 

A noteworthy activity that our manual classification of repositories brought forward is \textit{retrofitting}: the implementation of some existing language in Xtext.
We found 74 cases of repositories that can be classified as such.
In several cases, repositories included implementations of popular practical languages, e.g., GraphQL for graph database querying (\texttt{graphQL-xtext-grammar}), JSON Schema for schema definition (\texttt{JSON-Schema-to-internal-DSL}), and PlantUML, a notation for lightweight UML diagram creation (\texttt{plantuml-eclipse-xtext}).
The added value of an Xtext implementation for these languages is to benefit from the editing support offered by Xtext, e.g., automated code completion, syntax highlighting, and formatting.
A few cases were concerned with historical programming languages and could be seen as an enthusiastic effort (e.g., \texttt{Oberon-IDE}).

The majority of retrofitting cases was developed in a grammar-driven way, which is consequential: The concrete syntax in these cases is fixed and hence, it is natural to specify a grammar matching the existing syntax, thus retrofitting it.


\begin{tcolorbox}[colback=white!95!black, colframe=white!50!black, arc=3mm, left=0.5em, right=0.5em, top=0.5em, bottom=0.5em]
    \textbf{Results of RQ3: While the majority of Xtext-based language development projects involve the grammar-driven scenario, a nonnegligible number relies on the metamodel-driven one. We identify \textit{retrofitting}---creating an Xtext implementation of an existing language---as a noteworthy development activity.}
\end{tcolorbox}

\subsection{\reviseX{RQ4: How do Xtext-based languages on GitHub evolve over time?}}

\subsubsection{\reviseX{RQ4.1: How can evolution in these projects be characterized quantitatively?}}

To investigate how languages and their artifacts evolve, we focused again on the \revised{226} repositories classified as `language'.

\begin{table}[b]
    \centering
    \caption{Timespans of commits in repositories and  Xtext files.}
    \label{tab:time_span}
    \begin{threeparttable}
    \begin{tabular}{c lr l r}
      \toprule					
        ~ 	& \multicolumn{2}{c}{\textbf{Repo}} & \multicolumn{2}{c}{\textbf{Xtext}}\\
       \cmidrule(lr){2-3}\cmidrule(lr){4-5}
        Timespan (Days) & \#Repo  	& Perc. & \#Repo & Perc.\\
        \midrule
        >= 1000 	& 36    & 16.43\% & 18    & 8.22\%  \\       
        \midrule
        100 - 999 	& 55    & 25.11\% & 43    & 19.63\%  \\ 
        \midrule
        10 - 99		& 59    & 26.94\% & 44    & 20.09\% \\ 
        \midrule
        1 - 9       & 26    & 11.87\% & 29    & 13.24\% \\ 
        \midrule
        0           & 43    & 19.63\% & 85    & 38.81\% \\
      \bottomrule
    \end{tabular}
    \end{threeparttable}
\end{table}

\excl{ \begin{table*}[tb]
\centering
\caption{Ratio between time spans of repository activity and  time spans of grammar updates per repository.}
\label{tab:duration}
\begin{tabular}{c|c|c|c|c|c|c|c|c|c|c}
\hline
\textbf{Percentage} & \textbf{0\%-10\%} & \textbf{10\%-20\%} & \textbf{20\%-30\%} & \textbf{30\%-40\%} & \textbf{40\%-50\%} & \textbf{50\%-60\%} & \textbf{60\%-70\%} & \textbf{70\%-80\%} & \textbf{80\%-90\%} & \textbf{90\%-100\%} \\
\hline
\textbf{\#Repos}      & 96 & 8 & 7 & 4 & 7 & 3 & 8 & 10 & 10 & 69 \\
\hline
\end{tabular}
\end{table*} }

\excl{
\begin{table}[tb]
\centering
\caption{Average number of commits for textual instances in repositories that contain at least one Xtext file.}
\label{tab:aver_commit_ins}
\begin{tabular}{c|l|r}
\hline
\textbf{Avg. \#commits of instances} & \textbf{\#Repos} & \textbf{Percentage} \\
\hline
10 - 99      & 3  & 3.37\% \\
2 - 9        & 47 & 52.81\% \\
1            & 39 & 43.82\% \\
\hline
\end{tabular}
\end{table}

\begin{table}[tb]
\centering
\caption{Average number of commits for textual instances at different levels of average number of commits to grammar.}
\label{tab:xtext_ins_co_evolution}
\begin{tabular}{c|l|c}
\hline
\textbf{Avg. \#commits grammar} & \textbf{\#Repos} & \textbf{Avg. \#commits ins.} \\
\hline
1              & 25 & 1.1 \\
> 1 and <= 10  & 41 & 1.85 \\
> 10           & 23  & 4.78 \\
\hline
\end{tabular}
\end{table}
}

\begin{table}[tb]
\centering
\caption{Average change to number of lines and rules when comparing first and last commited version of Xtext grammar.}
\label{tab:xtext_line_change}
\begin{tabular}{c|l|r}
\hline
\textbf{Avg. Change to \#Lines} & \textbf{\#Repos} & \textbf{Percentage} \\
\hline
< 0          & 4 & 5.02\% \\
0            & 78 & 35.62\% \\
> 0 and <= 10       & 27   & 12.33\% \\
> 10 and <= 100     & 68   & 31.05\% \\
> 100 and <= 1000   & 33    & 15.07\% \\
> 1000              & 2     & 0.91\% \\
\hline
\end{tabular}
\label{tab:grammar_rule_change}
\begin{tabular}{c|l|r}
\hline
\textbf{Avg. Change to \#Rules} & \textbf{\#Repos} & \textbf{Percentage} \\
\hline
> 100           & 3     & 1.37\% \\
<= 100 and > 10 & 46    & 21\% \\
<= 10 and > 0   & 65   & 29.68\% \\
= 0             & 91   & 41.55\% \\
< 0             & 14   & 6.39\% \\
\hline
\end{tabular}
\end{table}

We were interested in the longevity of language projects and the activity around their included grammars.
To study this aspect, we focused on those \lang-classified repositories that contained a grammar, leading to 217 repositories. We observed a time span for updates of the repository and the  contained grammar. The time span is calculated as the difference in days between the first and the last commit. We obtained the results shown in Table~\ref{tab:time_span}.
The results show that \revised{43} repositories were never updated after they were created, while \revised{36} repositories had updates spanning more than 1,000 days, meaning that they were still maintained  after a substantial amount of time.
Considering the time span of updates to the grammar included in the projects, there are many (i.e. \revised{85}) repositories where Xtext files are never updated after they are initially created. There is also a nonnegligible number (i.e. \revised{18}) of repositories that still have records of updating Xtext files after a long time (no less than 1000 days).

We were further interested in how much Xtext grammars change over time.
To this end, we considered both the changes to the overall number of lines and the number of rules.
Starting with the number of lines,
for those 217 repositories that contain at least one Xtext file and are classified as \lang, we compared the first and the last committed version of the Xtext files in terms of their line counts, reporting the averages per repository in Table~\ref{tab:xtext_line_change}. 
We can see that the number of lines of text for Xtext grammar has not changed in \revised{35.62\%} of the repositories.
For those repositories that have changed, the vast majority of them contain Xtext files that have basically added the number of lines of text. 
Among them, the average increased line number of Xtext files in \revised{35} repositories by more than 100. 


We also analysed the changes in the number of grammar rules in the Xtext file, leading to the results shown in Table~\ref{tab:grammar_rule_change}. 
\revised{We can see that in more than 40\% of the repositories, the number of grammar rules in the Xtext grammar has not changed.}
Part of the reason is that \revised{66} repositories have not updated their Xtext files since they were initially committed. 
For those repositories where the number of rules of the Xtext grammar changed, in most of them  the number of  rules increased over time. In particular, the average number of added grammar rules in the Xtext grammar contained in \revised{47} repositories exceeds ten.

Our dataset \cite{dataset} contains additional metadata quantifying evolution activities, specifically, the change frequency of grammars, metamodels, and instances per project.
While a detailed analysis is outside the scope of this paper, we observe the following trends:
As one would expect, grammars are  updated more often than meta-models.
However, the difference in update frequency is more pronounced in the grammar-driven workflow than in the metamodel-driven one, which might suggest that users in this workflow spend more time polishing concrete syntax aspects. 
Instances are more likely to be updated in repositories with more Xtext grammar updates. 


\subsubsection{\reviseX{RQ4.2: How common are different types of changes during the evolution of grammars?}}

\reviseX{As mentioned in section 5.2.1 that} 217 repositories classified as ``Language'' that contain at least one xtext file. In Step 7, we obtained all commits of Ecore files, Xtext files, and instance files from these 217 repositories, totaling 4817. After manual review, we excluded 24 commits about xml files that were incorrectly identified as instance files. Therefore, the total number of commits is 4793.
We counted the number of days between each commit and the previous commit of the same file. 
We found that in 192 repositories, the commits to the instance files were never more than 5 days away from the previous commit to the same file. In 115 repositories, the commits to the xtext files were never more than 5 days away from the previous commit to the same file.
There is an overlap of 111 repositories in which both, grammars and instances, only show incremental changes with 5 or fewer days between commits.
Thus, there are only 21 repositories classified as ``Language'' where both, grammar and instance files, exist and show changes that are potentially not just incremental development.
\reviseX{
In addition, we found that there were 3809 commits of files that had previously been committed within five days. This hints that DSLs, just like other software, are not carefully designed on paper and then implemented once, but are built in iterations, implying some form of rapid prototyping of the grammar development.}

\reviseX{We also found that there are 438 commits of xtext, \reviseX{Ecore,} or instance files that are more than 30 days away from the previous commit of the same file.\footnote{\reviseX{Commits including two or more xtext, Ecore, or instance files that have previously been committed more than 30 days ago are counted two or more times accordingly.}}  These commits come from 78 repositories.}

\begin{figure}[htbp]
\centering
    \includegraphics[scale=0.7]{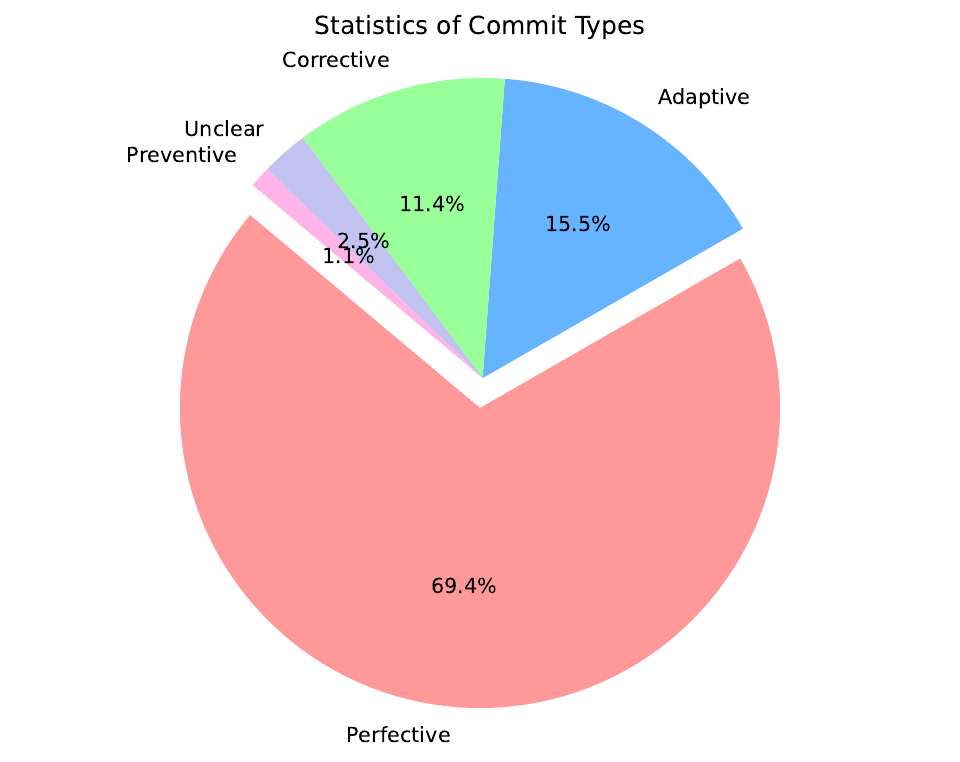}
\caption{Statistics of the number of different commit types, \reviseX{i.e., adaptive, perfective, corrective, and preventive commit.}}
\label{fig:commit_types}
\end{figure}

\reviseX{We classified these 438 commits according to the types described in Step 7. We found that about two-thirds of these commits (304 in total) are ``perfective commits''. The number of ``adaptive'' and ``corrective" commits is also not small, with 68 and 50 respectively. In addition, there are five commits, only, that belong to the ``preventive'' type. The proportion of these different types of commits to the total number is shown in Figure~\ref{fig:commit_types}.}
\reviseX{
The high number of perfective changes indicates that there is an inherent (functional) need for these languages to evolve. Thus, the changes seen are not just for maintenance, which would be rather associated with corrective and adaptive changes. This indicates that changing requirements on the language are the main driver of DSL evolution. 
}

\subsubsection{\reviseX{RQ4.3: How do textual instances co-evolve with grammar in real projects?}}

To answer RQ4.3, we followed step 7 of our methodology, i.e., for those commits files that were more than 30 days away from the previous commit of the same file, we determined whether there was another commit in the same repository that was no more than 5 days away from it but committed a different type of file. 
For example, in the repository named ``gen-angular-grammar'' with login ``jhonatan89'', the meta-model file ``Generator.ecore'' was committed on February 2, 2018, we checked if there was another file that was of a different type than the file ``Generator.ecore'' and it was committed within five days around February 2, 2018.
We filled in ``YES/NO'' in a new column. \reviseX{We found that there were 188 such commits, and they came from 39 repositories.} 

We checked these 188 commits and the 39 repositories where they were located, and found that 36 of them had a 
\reviseX{scenario}
of co-change, i.e., two language artifacts of different file types were changed at the same time or changed one after another within five days, and more than 30 days later, the two files were changed again at the same time or changed one after another within five days. For example, still in the repository ``gen-angular-grammar'', the metamodel file ``Generator.ecore'' and the xtext file ``Generator.xtext'' evolved simultaneously on November 30, 2017, and they evolved simultaneously again on February 2, 2018. Among these 36 repositories, 33 of them 
\reviseX{had metamodels and Xtext files that changed simultaneously or within five days of each other, eleven of them had Xtext files and instances that changed simultaneously or within five days of each other, and eight of them involved both cases.}

By combining these observations with the analysis of commit types, we found that among these 188 commits, there are 151 ``perfective'' commits, 22 ``adaptive'' commits, ten ``corrective'' commits, two ``preventive'' commits, and another three commits are ``unclear''. Therefore, compared with the proportion of each commit type in the previous section, the proportion of ``perfective'' commits here has increased, while the proportion of almost all other commits has decreased.

\begin{tcolorbox}[colback=white!95!black, colframe=white!50!black, arc=3mm, left=0.5em, right=0.5em, top=0.5em, bottom=0.5em]
    \textbf{Results of RQ4:}
    We find a spectrum of project and language development lifespans of Xtext projects on GitHub, with a non-negligible part (8\%) of grammars still being updated 1000 days after the initial commit.
    The amount of change performed in individual languages can be significant, with more than 10 rules being added in 22\% of all languages.
    We also extracted 4,793 commits from 217 repositories and found that in about half of the repositories, no commit of Xtext files and instance files was more than five days away from the previous commit. Moreover, most cases of evolution were classified as ``perfective'' changes, indicating changing requirements on the DSLs. We also found co-evolution of files of different file types (instance, grammars, and meta-models) in 36 repositories.
\end{tcolorbox}

\section{Discussion}
\label{sec:discussion}
We will discuss our results from three perspectives: the need for approaches to support co-evolution in textual DSLs, the need for a better understanding of DSL development in practice, and threats to the validity of our research.

\subsection{Need for approaches for co-evolution in textual DSL evolution contexts} 
\label{sec:discuss-coevol}
Our analysis in RQ3 reveals the existence of DSL evolution projects that are maintained over several years and involve significant changes to the language over time.
In such projects, challenges arise from the synchronized evolution of all involved artifacts, including grammars, meta-models, instances, and workflow files.
In a general MDE context, synchronized evolution is a well-studied area,  termed  \textit{co-evolution} or \textit{coupled evolution}.
A plethora of existing work \cite{garcia2012model,khelladi2017exploratory,kusel2015consistent,vaupel2015agile,kessentini2020interactive,di2023modeling,di2013methodological}  provides foundational approaches for managing co-evolution between meta-models and other artifacts, typically models and transformations.
However, there remains a significant gap in tool support and methodologies specifically tailored to the co-evolution of artifacts involved in textual DSLs developed with frameworks like Xtext, with their focus on both meta-models and grammars.

\reviseX{Our observations hint on a number of requirements that need to be taken into account when approaching the co-evolution problem for textual DSLs:
\begin{itemize}
 \item Change and specifically evolution of DSLs is not just about maintenance activities, as witnessed by the majority share of ``perfective'' changes in cases where grammars and instances change after more than 30 days (see Figure \ref{fig:commit_types}). Thus, DSLs need to be seen as software systems that will undergo change in their functionality and therefore also be supported as such.
 \item Rapid iterations of the DSL happen. We observe for 106 of all repositories that are classified as languages changes in instances, meta-models, and grammars within less than 5 days. DSL frameworks need to support such rapid iterations.
 \item Both grammar-driven and meta-model driven development happens and is subject to evolution. Frameworks need to support both evolution cases.
 \item Co-evolution between instances and grammar happens and our numbers likely underestimate the frequency. In 39 out of 275 repositories with grammars and instances we observed that both grammars and instances change together after a break of more than 30 days, hinting on co-evolution. However, since instances in the DSLs repository are likely just example and serve testing and documentation purposes, this number does not include all cases of co-evolution needed for productive instances of the DSLs, which are not seen in these repositories.  
\end{itemize}}

To address some of the complexity in the meta-model-driven scenario (RQ2), one could rely on principles from the grammarware sphere and support an operator-based approach to grammar evolution   \cite{zaytsev2014negotiated}.
In our recent work, we follow up on this idea to 
 automate parts of the synchronization of the grammar after meta-model changes \cite{zhang2023automated,zhang4379232supporting,zhang2023rapid,zhang2023towards}.
Another open challenge is grammar-instance co-evolution, which could benefit from the available foundational approaches for metamodel-model co-evolution, but needs to deal with concrete syntax aspects of grammars.
This problem has not been addressed yet in the technical space of Xtext.
Lämmel's LAL approach \cite{lammel2016coupled} could be useful for validating a solution that addresses it.

Future research should focus on integrated approaches and tools that facilitate the simultaneous evolution of all related DSL artifacts.
The comprehensive dataset compiled from our study, particularly the 196 cases where all crucial artifacts such as grammar, metamodel, and instances are available, presents a valuable resource for developing new co-evolution approaches, by supporting their design, testing and evaluation.

\subsection{\reviseX{Need for a better understanding of DSL development in practice}}
\label{sec:discuss-underst}
\reviseX{Several observations show that we need a better understanding of how DSL development looks like in practice in order to better support it. This concerns practices such as testing and documentation, versioning support, and the accessibility to domain experts outside computer science.}

\subsubsection{\reviseX{What testing and documentation practices are used for the DSLs' development?}}
\reviseX{While some of the studied repositories that are classified as languages include instances, it is only a minority. Even among those repositories with instances the majority does not include a sufficient variety of instances to cover all grammar rules. Thus, in only 18 cases, we found instance examples covering the complete grammar of the DSL. 
This leads to the question how these DSLs are tested and documented. 
Instances central to testing a language. Without instances, it is not possible to evaluate whether the created language is usable.
Similarly, language documentation, e.g., tutorials, rely on instance example. The most famous being variations of ``Hello World'', that are used to introduce language users to basic language concepts and allow them to get started learning a language.
Thus, it is unclear how testing and documentation is done for most of the languages in the found repositories.
On the other hand, it is possible that the documentation of the DSL is stored outside of the DSL's main repository. 
This would be explainable with the architectural setup of Xtext, in which language developers use separate Eclipse instances and associated workspaces to edit and to test the language definition.
Hence, language definition artifacts (grammars and meta-models) are stored in different workspaces than example cases, and language engineers need to invest dedicated effort in order to make available example cases together with the language definitions.
Still, this hints that there is a difference to the practices in other open source software systems, where tests are often stored in the same repository as the code and documentation is at least linked from the repository. 
Future work is required to investigate DSL testing and documentation practices in OSS. 
}

\subsubsection{\reviseX{Is there missing support to version Xtext projects?}}
\reviseX{In RQ1, we found that out of the 280 repositories that did not contain Xtext files, a number of them contained other MDE artifacts. Specifically, 9 repositories contain at least one Ecore metamodel but no Xtext files. Possible explanations for these cases are: First, the repository owners were aiming at a meta-model-driven scenario, but did not complete the definition of the grammar. Second, the repository owners created the Xtext files from an Ecore metamodel,  and generated them into a different directory that was not included in the repository. Similarly,  three of these repositories contain MWE files and two contain instance files, however, these repositories are often incomplete, as can be discovered by importing them into Eclipse, which leads to error messages.
This observation goes hand-in-hand with the observation above that test instances require other eclipse instances and, thus, other workspaces, which are likely to not be versioned together with the DSL itself. 
Future work needs to investigate whether the architecture of eclipse/Xtext makes it more difficult to access proper version management and establish the requirements for more suitable version management systems to support DSL development.
}

\subsubsection{\reviseX{Is Xtext only used for DSLs built for software developers?}}
\reviseX{The distribution of DSL domains reveals a strong focus on domains where the expected user are computer scientists themselves. 
Only surprisingly few of the DSLs are from other domains, e.g., Healthcare and Life Sciences, or DSLs for development of questionnaires.
This is remarkable, as one purpose of DSLs is to make programming accessible to domain experts that are not classical programmers.
The results could be explain in two ways:
It is of course possible that these results are representative of how DSLs are distributed. In this case, it would be worthwhile investigating why DSLs are not used or worth using in other domains.
However, more likely, the results are due to our focus on DSLs built with Xtext.
Thus, it seems that DSLs for domains outside computer science are less likely build with Xtext. The question is whether the same holds for other DSL frameworks, like Langium \cite{langium}, or whether those DSLs are mostly built without the use of language frameworks. This would imply that these DSLs are potentially even more vulnerable to issues like language evolution, a topic addressed in research for DSLs build on concepts such as MOF-like meta-models \cite{hebig2016approaches}. 
Future work needs to be done to answer the question how DSLs for non-computer-science domains are built and investigate the consequences of that.
We also need to ask ourselves as a community whether the frameworks we offer are accessible to researchers and engineers from other domains.
}

\subsubsection{\reviseX{What is the compliance relationship between artifacts in practice?}} 
\reviseX{Our comprehensive dataset paves the way for future research on language development and evolution, especially for detailed studies of language evolution extended on RQ4. In this paper, we have conducted preliminary research on the co-changes of different language artifacts in the same repository, but have not yet analyzed the actual compliance relationship between these language artifacts. I.e., for example, we found many cases where Ecore files and Xtext files co-evolve in the same repository, but it is not clear whether the Xtext file complies with the Ecore file and vice versa. Finding repositories with such scenarios of co-changes of different artifacts allows us to focus future work on a smaller set of repositories. In addition, this repository collection can help us find suitable case languages more quickly for developing and studying tools for grammar and instance co-evolution, ultimately fostering more robust and adaptable languages.}

\subsection{Threats to validity}
\label{sec:threads}
\paragraph{Threats to internal validity} arise from us relying on the GitHub API.
Since we cannot access the implementation of the GitHub API, we cannot verify \textit{completeness}: the implementation could be inexact, which could lead to repositories not being captured by our query.
As a safeguard, we checked whether a total of three expected repositories personally known to us appeared in our results, which was the case. 
Nevertheless, anecdotally, a colleague to whom we made available our dataset informed us about a project that was not part of it.
Still, our findings that arise from a substantial number of cases and highlight the existence of understudied phenomena---meta-model-based evolution, retrofitting, long-living Xtext projects--do not require completeness to be valid.

\excl{As one example, our strategy for identifying textual instances in RQ_instance relies on reading the file extensions for instances from  MWE files associated with grammars and then retrieving the files with these extensions to obtain the set of textual instances.
To check whether this strategy leads to accurate results, we checked for ten randomly sampled case repositories  whether the files identified as  instances for these repositories all indeed adhere to the Xtext grammar of the same repository and found that this applies.
Hence, we can conclude that our strategy is accurate in these cases.
Yet, it is possible that a small minority of the files retrieved that way might not  be actual instances.}

Furthermore, repositories can be duplicates of each other, which might bias the results.
For our \lang-classified repositories, which we investigate in RQ2 and RQ3, we checked that this is not the case and can exclude duplicates.
For the \experimental category, which generally does not make any statements about the quality about the included repositories, anecdotally, some repositories have the same contents as others, potentially arising from having followed the same tutorial with the basic Xtext examples.

\reviseX{In addition, in Step 5, we manually judged the domain of the repositories and the categories they belonged to. Potential bias posed an internal threat, so for each data, we used another person's review to reduce the bias of manual judgment. Similarly, we manually judged the type of commits in Step 7, which may also be biased. We had a second person to take a sample review to reduce bias.}

%

\paragraph{Threat to external validity.} Considering external validity, our scope is restricted to GitHub and Xtext, both being particularly popular and widely used  technologies in their respective communities. 
There is a larger variety of existing  language workbenches \cite{erdweg2015evaluating}, not all of which might equally benefit from our findings.
The results from our study could be transferable to other workbenches that use a similar strategy for separating abstract and concrete syntax specification like Xtext, such as textX and langium.
Yet, transferring our results to language workbenches that employ an entirely different paradigm (e.g., in the case of MPS,  projectional editing) might be infeasible.




\paragraph{Threats to construct validity.} Considering construct validity, \excl{from our use of the GitHub API search, which we used to identify Xtext-related repositories.
As we described in the description of the results for RQ1, a substantial number of cases (11\% of all considered repositories) were  related  to Xtext only due to their name.
Still, our manual classification of all repositories allowed us to identify these cases  and  avoid possible confusions when using them.
Furthermore,
our classification of repositories gives rise to threats related to subjectivity and inaccurate information.}to mitigate the impact of subjectivity on our classification, we extensively discussed the criteria and problematic cases and eventually found consensus for all of them.
Our classification further relies on  documentation provided by the repository owners, which might not always be accurate or complete, leading two consequences:
First, we did not investigate whether repositories were from industry or academia, which generally was not possible to tell from the documentation.
Second, some of our \textit{language}-classified repositories, in particular, among those classified as  \textit{retrofitting}, might  be exclusively intended for self-teaching or demonstration purposes, but this context is unavailable to us.
Users of our dataset are advised to use it in a way that makes sure that their assumptions are met, for example, taking into account our collected change history meta-information to identify cases with a rich evolution history.

\reviseX{In addition, a threat to construct validity exists in our approach to identifying DSL evolution through commit time intervals. We use time intervals between commits (>30 days for evolution, < five days for incremental changes) as a proxy for identifying evolutionary changes. However, this measurement approach may not fully capture the actual nature of DSL evolution. Significant evolutionary changes could occur within short time intervals (< five days), leading to false negatives. Conversely, long intervals between commits (>30 days) might not necessarily represent true evolutionary changes, potentially resulting in false positives. This threatens the construct validity of our study as the temporal proximity of commits may not always accurately reflect the nature and significance of the changes made to the DSL. To address this threat in our future work, we plan to combine multiple evolution indicators beyond commit intervals, including qualitative analysis of change types, expert reviews of the modifications, and in-depth case studies to establish more robust evolution identification criteria.}

\section{Conclusion}
\label{sec:conclusion}
\reviseX{In our analysis of 1002 GitHub repositories, we found that about one fifth of them contain fully developed DSLs. We analyzed trends in in specific applications domains, and found out that specific domains, such as data management and databases, are maintained more frequently and for a loner time.
We also found that} the majority of 
\reviseX{those} languages are developed following a grammar-driven approach, albeit a notable number adopt a metamodel-driven approach. 
We observed long-running projects with numerous changes, highlighting the diversity in DSL development and the need for tools and methodologies that can accommodate different scenarios.
The trend of retrofitting existing languages in Xtext showcases its flexibility beyond creating new DSLs.
\reviseX{By investigating software evolution aspects, we found that the development lifecycle of DSLs varies, but in most DSL development projects, updates to grammar definitions and example instances are very frequent, and most of the evolution activities can be classified as ``perfective'' changes.}
Our dataset supports further research into DSL evolution and the development of methods to facilitate this evolution.

Future work will investigate our dataset,  
\reviseX{including discovering more insights into DSL testing and documentation practices and studying detailed evolution patterns,}
particularly in Xtext constructs and code artifacts. We also aim to develop better methods for synchronizing artifacts in evolution scenarios, especially co-evolution between a grammar and its instances, a gap not addressed by current solutions.
\section{COMPLIANCE WITH ETHICAL STANDARDS}
\textbf{Conflict of Interest:} The authors declare that they have no conflict of interest.\\
\textbf{Funding:} This research received no external funding.\\
\textbf{Ethical approval:} This study does not involve any human participants or animals.\\
\textbf{Informed consent:} Not applicable.\\
\textbf{Author Contributions (if more than one author):} Author One participated in the methodological design, data collection, results analysis, and writing, and developed various scripts required to implement the data collection process. Author Two proposed the topic of this article and participated in the methodological design, data collection, results analysis, and writing. Author Three participated in data collection and writing.\\
\textbf{Data availability statement:} The data is fully open in~\cite{dataset}.

\balance 
\bibliographystyle{spbasic}
\bibliography{main}

\end{document}